\documentclass[letterpaper, 10 pt, conference]{ieeeconf}  

\IEEEoverridecommandlockouts                              
\usepackage[utf8]{inputenc}
\usepackage[T1]{fontenc}

\usepackage{graphics} 
\usepackage{epsfig} 
\usepackage{mathptmx} 
\usepackage{mathptmx} 
\usepackage{amsmath} 
\usepackage{amssymb}  

\usepackage{textcomp}
\usepackage{txfonts}

\newcommand\Tstrut{\rule{0pt}{2.0ex}}         
\newcommand\Bstrut{\rule[-0.9ex]{0pt}{0pt}}   

\usepackage{color}
\usepackage[ruled,vlined,algo2e]{algorithm2e}
\SetKwComment{Comment}{$\triangleright$\ }{}

\usepackage{pifont}
\newcommand{\cmark}{\ding{51}}%
\newcommand{\xmark}{\ding{55}}%

\usepackage{multirow}
\usepackage[colorlinks=true, allcolors=blue]{hyperref}

\usepackage{fancyhdr}
\usepackage{kantlipsum}
\fancypagestyle{plain}{
  \fancyhf{}
  \fancyhead[L]{}     
  \fancyfoot[L]{}

}
\usepackage{eso-pic}

\title{\LARGE \bf Limited View Tomographic Reconstruction Using a Deep Recurrent Framework with Residual Dense Spatial-Channel Attention Network and Sinogram Consistency}

\author{Bo Zhou, S. Kevin Zhou, James S. Duncan, Chi Liu
\thanks{B. Zhou is with the Department of Biomedical Engineering, Yale University, New Haven, CT, 06511, USA.}
\thanks{J. S. Duncan and C. Liu are with the Department of Biomedical Engineering and the Department of Radiology and Biomedical Imaging, Yale University, New Haven, CT, 06511, USA.}
\thanks{S. K. Zhou is with the Institute of Computing Technology, Chinese Academy of Sciences, Beijing, 100190, China.
Corresponding email: bo.zhou@yale.edu and chi.liu@yale.edu}
}

\begin{document}

\maketitle

\AddToShipoutPictureBG*{%
  \AtPageUpperLeft{%
    \setlength\unitlength{1in}%
    \hspace*{\dimexpr0.5\paperwidth\relax}
    \makebox(0,-0.75)[c]{\scriptsize THIS WORK HAS BEEN SUBMITTED TO THE IEEE FOR POSSIBLE PUBLICATION. COPYRIGHT MAY BE TRANSFERRED WITHOUT NOTICE.}%
}}

\begin{abstract}
Limited view tomographic reconstruction aims to reconstruct a tomographic image from a limited number of sinogram or projection views arising from sparse view or limited angle acquisitions that reduce radiation dose or shorten scanning time. However, such a reconstruction suffers from high noise and severe artifacts due to the incompleteness of sinogram. To derive quality reconstruction, previous state-of-the-art methods use UNet-like neural architectures to directly predict the full view reconstruction from limited view data; but these methods leave the deep network architecture issue largely intact and cannot guarantee the consistency between the sinogram of the reconstructed image and the acquired sinogram, leading to a non-ideal reconstruction. In this work, we propose a novel recurrent reconstruction framework that stacks the same block multiple times. The recurrent block consists of a custom-designed residual dense spatial-channel attention network. Further, we develop a sinogram consistency layer interleaved in our recurrent framework in order to ensure that the sampled sinogram is consistent with the sinogram of the intermediate outputs of the recurrent blocks. We evaluate our methods on two datasets. Our experimental results on AAPM Low Dose CT Grand Challenge datasets demonstrate that our algorithm achieves a consistent and significant improvement over the existing state-of-the-art neural methods on both limited angle reconstruction (over 5dB better in terms of PSNR) and sparse view reconstruction (about 4dB better in term of PSNR). In addition, our experimental results on Deep Lesion datasets demonstrate that our method is able to generate high-quality reconstruction for 8 major lesion types.
\end{abstract}


\section{INTRODUCTION}
\begin{figure}[htb!]
\centering
\includegraphics[width=0.47\textwidth]{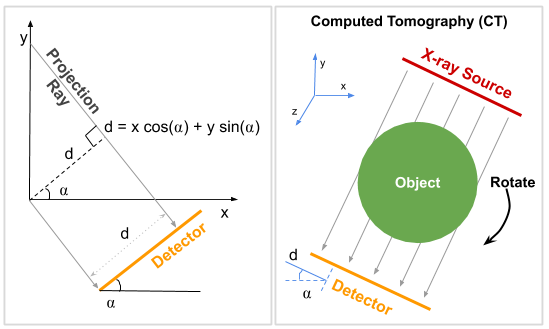}
\caption{Illustrations of CT scanners and its projection geometry. CT uses an external rotating x-ray source for projection. The sinograms of projection measurement are reconstructed into tomography images.}
\label{fig:ill}
\end{figure}

Tomography imaging is a non-invasive projection-based imaging technique that visualizes an object's internal structures and hence finds wide applications in healthcare, security, and industrial settings \cite{anirudh2018lose,de2014industrial,zhou2019limited}. 
In healthcare, tomography imaging techniques such as medical Computed Tomography (CT) based on x-ray projections, Positron Emission Tomography (PET), and Single-photon Emission Computed Tomography (SPECT) based on gamma-ray projections are indispensable imaging modalities for disease diagnosis and treatment planning. In the traditional CT setting, one assumes access to the measurements that are collected from a full range of view angles of an object, i.e., $\alpha \in [0, 180^{\circ}]$. To reduce radiation dose and speed up acquisition, recently it is of increasing interest to develop methods that can recover images when a portion of the sinogram/projection views is missing, namely limited view tomographic reconstruction. There are two notable sub-problems: limited angle (LA) reconstruction, i.e., when $\alpha \in [0, \alpha_{max}]$ with $\alpha_{max} < 180^{\circ}$ for equivalent parallel beam geometry, and sparse view (SV) reconstruction with a view interval larger than normal. Both LA and SV acquisitions can efficiently reduce radiation dose. Using LA acquisition, the scan time can also be drastically reduced by restricting the physical movement of the scan arc. Note that fast acquisition or high temporal resolution is paramount; even a slightly longer scan time can lead to appreciable motion blur and and artifact in the image \cite{cho2013motion,mohan2015timbir}. The illustrations of projection geometry for generating SV or LA sinogram are shown in Figure \ref{fig:ill}.

There are two major factors, namely reconstruction quality and speed, that need to be properly considered in designing a tomographic reconstruction algorithm. Currently, Filtered Back Projection (FBP) is widely used as the standard algorithm as it can reconstruct a high-quality image with a fast speed, following an analytical solution. However, FBP assumes the access to the measurements that are collected from a full range of views of an object, i.e., $\alpha \in [0, 180^{\circ}]$. Reconstruction using FBP in both LA and SV conditions are highly ill-posed, yielding non-ideal image quality with severe artifacts and high noise. Previous algorithms for tomographic reconstruction under limited view conditions can be classified into two general categories: model-based iterative reconstruction (MBIR) and deep learning based reconstruction (DLR). MBIR can generate images with diagnostic quality by minimizing the predefined image domain regularizers and the sampled sinogram inconsistency in an iterative fashion. Common choices of the regularizer include total variation \cite{chambolle1997image}, dictionary learning \cite{xu2012low}, and nonlocal patches \cite{zhang2013iterative}. However, MBIR methods are computationally heavy and time-consuming since they rely on repetitive forward and backward projections. Moreover, using regularization solely based on prior assumptions requires careful hyper-parameter tuning and tends to bias the reconstruction results, especially when under-sampling rate is high. 

Recently, deep learning techniques, such as convolutional neural networks (CNNs), have been widely adapted in tomography and demonstrated promising reconstruction performance \cite{wang2018image}. Combining MBIR with deep learning, Gupta et al. \cite{gupta2018cnn} and We et al. \cite{wu2017iterative} first proposed to model regularizer in MBIR frameworks with CNNs and Autoencoders. Adler et el. \cite{adler2018learned} unfolded the optimization procedure of MBIR to an N-stage network to balance the trade off between reconstruction and speed. Although improved over traditional MBIR methods, they still suffer from high computational cost with iterative procedures.
As an alternative, DLR is often formulated as image post-processing. Jin et al. \cite{jin2017deep} and Chen et al. \cite{chen2017low} proposed to use UNet \cite{ronneberger2015u} and Residual UNet to post-process the noise/artifacts in the sparse-view CT. In \cite{yang2018low} and \cite{liao2018adversarial}, adversarial loss and perceptual loss were used to reinforce the network's learning. Later, Zhang et al. \cite{zhang2018sparse} and Han et al. \cite{han2018framing} proposed to incorporate dense block and wavelet decomposition into UNet for more robust feature learning for reconstruction. Direct sinogram inversion and sinogram completion strategies were also proposed. Anurudh et al. \cite{anirudh2018lose} proposed to directly encode limited angle sinogram and decode as reconstructed image. Lee et al. \cite{lee2018deep} found that synthesizing complete sinogram from sparse view sinogram and then using FBP can also reconstruct high-quality image. Although these methods can be easily applied to raw sinograms or corresponding FBP reconstructed images with relatively low computational cost and low design complexities, they either only applied on image domain that remove artifacts in already reconstructed image or synthesizing complete sinogram from sparse one, and cannot guarantee the sampled sinogram data are preserved. Note that the sampled sinogram data are the original sources that should be kept as identical as possible before and after reconstruction to ensure the high fidelity of reconstructed content. While there are recent ideas of replacing the already-sampled sinogram to the predicted sinogram during only the test stage \cite{anirudh2018lose,huang2019data}, this strategy does not guarantee the consistency between the already-sampled sinogram and the predicted sinogram, which may further degrade the final reconstruction. On a different note, the network design issue is highly under-explored as a research topic and still limited to UNet-based or auto-encoder architectures \cite{jin2017deep,chen2017low,han2018framing,yang2018low,liao2018adversarial,lee2018deep,kofler2018u}. In addition, none of previous works have evaluated the performance under both LA and SV scenarios, and reconstruction evaluation on CT scan with pathological finding are barely performed. While a k-space data consistency layer for MRI fast reconstruction is proposed in~\cite{schlemper2017deep,zhou2020dudornet}, sinogram consistency layer has not been systematically studied in tomographic reconstruction. 

To tackle these limitations, we propose a deep recurrent reconstruction framework for high-quality tomographic reconstruction under limited view conditions, which stacks the same block multiple times. In each recurrent reconstruction block, we design a residual dense spatial-channel attention network (RedSCAN) and use it as our backbone network. We call our recurrent framework as R$^2$edSCAN. To ensure the reconstruction prediction's sinogram consistency, we also design a Sinogram Consistency Layer (SCL) interleaved in our recurrent framework that ensures the sampled sinogram is consistent with the sinogram of the intermediate outputs of the recurrent blocks, thus improving the fidelity of the final reconstruction. Experiments on limited angle and spare view scans on different anatomical areas and images with lesions demonstrate that our R$^2$edSCAN framework, consisting of RedSCAN and SCL, generates a superior tomographic reconstruction as compared to previous state-of-the-art methods.

\begin{figure*}[htb!]
\centering
\includegraphics[width=0.98\textwidth]{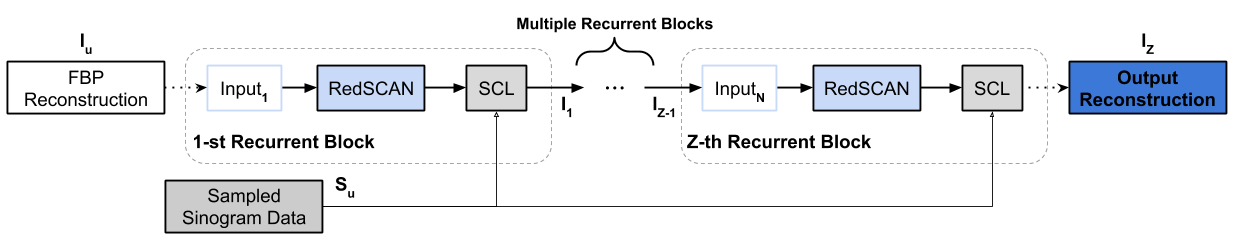}
\caption{The overall pipeline of our R$^2$edSCAN framework. Each recurrent block consists of a RedSCAN (blue) and a SCL (gray).}
\label{fig:pipeline}
\end{figure*}

\section{THEORY AND PROBLEM FORMULATION}
Without loss of generality, the reconstruction discussed in this work refers to parallel-beam imaging geometry, and can be adapted to other projection geometries by rebinning their sinogram into parallel-beam geometry first. As illustrated in Figure \ref{fig:ill}(d), the parallel-beam forward projection for an arbitrary 2D object $\rho (x,y)$ can be written as:
\begin{equation}
    m(d, \alpha) = \int \int \rho (x,y) \delta(x cos \alpha + y sin \alpha -d) dx dy
\end{equation}
where $m(d, \alpha)$ is the projection measurement at detector position $d$ and rotation angle $\alpha$. $\delta(\cdot)$ is the standard Dirac delta function. At each angle of the object, the measurements at the detector is the line integration of the attenuation coefficient values along the projection path. Measurements at different angles construct our sinogram data. 

Let $I \in \mathcal{C}^{N}$ represent a 2D tomography image containing a 2D object of $\rho (x,y)$ with a size of $N=N_x N_y$, and $S \in \mathcal{C}^M$ represent its full-view sinogram with $M$ projection views. Our problem is to reconstruct $I$ from $S_u \in \mathcal{C}^{M_u}$ $(M_u \ll M)$, where $S_u$ is the under-sampled sinogram of limited views. Here, sinogram data is only measured for lines corresponding to a subset $\Omega \subset \mathcal{A} \overset{\Delta}{=} \{1,\cdots,M\}$, where $\mathcal{A}$ is the full projection set. Denoting $\varmathbb{F}$ and $\varmathbb{F}_u$ as the full-view and limited-view discretized forward projection operators, the full-view sinogram $S$ and limited-view sinogram $S_u$ are obtained via $S=\varmathbb{F} I$ and $S_u=\varmathbb{F}_u I$, respectively. The pseudo-inverse of full-view forward projection operators, denoted as $\varmathbb{B}$, can be solved by FBP. While FBP provides stable numerical implementation of $\varmathbb{B}$ for $S$, applying $\varmathbb{B}$ to $S_u$ in the limited view conditions yields reconstructed $I_u$ with severe artifacts and high noise level.

Previous works of MBIR propose to solve $I$ by minimizing $\mathcal{T}(I) + \lambda || \varmathbb{F}_u I - S_u ||^n_n$, where $\mathcal{T}$ is the regularizer and $|| \cdot ||^n_n$ is the sinogram consistency constraint \cite{chambolle1997image,zhang2016low}. Previous deep learning-based, post-processing methods utilize deep networks, denoted as $\mathcal{P}$ with parameters $\theta$, to estimate the full-view reconstructed image $I=\mathcal{P}(I_u; \theta)$ by training $\mathcal{P}$ on $(I_u, I)$ pairs. However, these methods only consider a subsequent regularization of the initial solution $I_u$ similar to the functionality of $\mathcal{T}(\cdot)$ in MBIR, and omit the sinogram consistency constraint of $|| \varmathbb{F}_u I - S_u ||^n_n$. In this work, we formulate a deep model that directly predicts reconstructed image from sinogram by jointly optimizing both terms:
\begin{equation} \label{eq:general}
\underset{\theta}{\arg\min} ( || I - \mathcal{P}(I_u; \theta) ||^2_2 + \lambda || \varmathbb{F}_u \mathcal{P}(I_u;\theta) - S_u ||^2_2 ) ,
\end{equation}
where the second term is the sinogram consistency constraint achieved by our SCL. The above target function aims to estimate full-view reconstruction while discouraging deviation from the already measured sinogram from projection subset $\Omega$. Directly optimizing the above target function with traditional one-step CNN is challenging, due to its high computational complexity, over-fitting, and local optima issues. Thus, we propose to use a recurrent reconstruction framework with SCL that optimizes $\theta$ with a fixed number of iterations. The overall pipeline is shown in Figure \ref{fig:pipeline} and its details are discussed in the following sections.

\section{METHODS}
The general pipeline of our R$^2$edSCAN is shown in Figure \ref{fig:pipeline}. It consists of three parts: 1) RedSCAN with residual dense spatial-channel attention blocks, 2) SCL for regularization, and 3) recurrent reconstruction framework enabled by them. In the following section, we first present our RedSCAN with spatial-channel attention module in \ref{section:rdscan}. Then, we elaborate our SCL in \ref{section:scl}. Finally, we conclude the overall workflow and loss in \ref{section:rrf}.

\subsection{Residual Dense Spatial-Channel Attention Network}
\label{section:rdscan}
\begin{figure*}[htb!]
\centering
\includegraphics[width=0.89\textwidth]{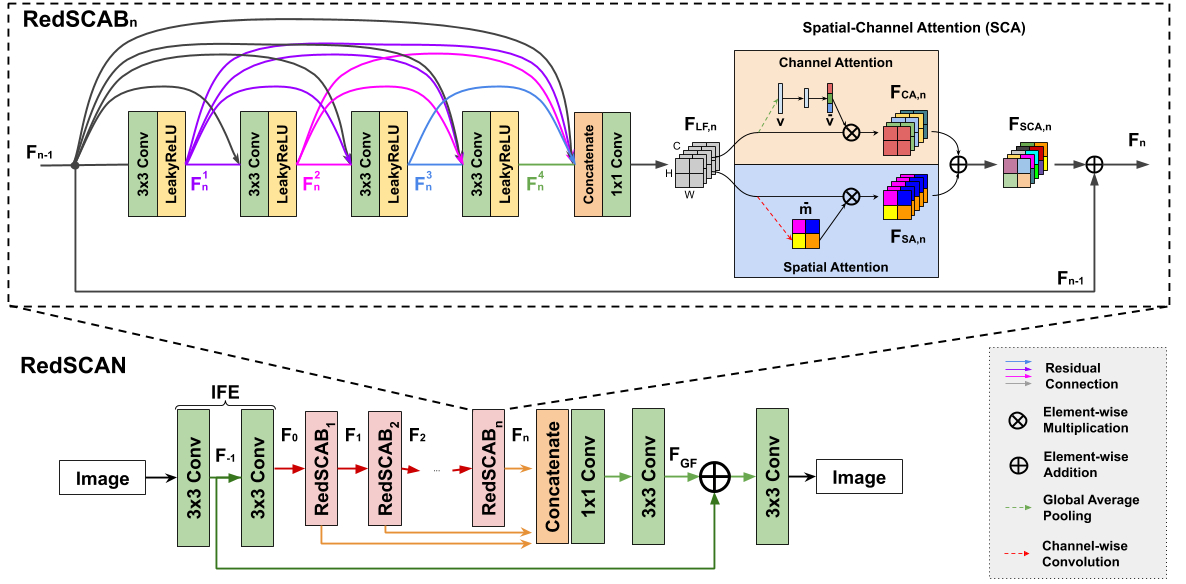}
\caption{The architecture of our Residual Dense Spatial-Channel Attention Network (RedSCAN), which are used in both the recurrent image reconstruction blocks in Figure \ref{fig:pipeline}.}
\label{fig:rdscan}
\end{figure*}

Our RedSCAN consists of three key components, including initial feature extraction (IFE) using two $3 \times 3$ convolution layers, multiple Residual Dense Spatial-Channel Attention Block (RedSCAB) followed by global feature fusion, and global residual learning. The network architecture is demonstrated in Figure \ref{fig:rdscan}. 

Let $\mathcal{P}_{IFE_{1}}$ and $\mathcal{P}_{IFE_{2}}$ be the first and second convolutional operations in IFE, we first extract $F_{-1} = \mathcal{P}_{IFE_{1}} (I_u)$ for global residual learning, and $F_{0} = \mathcal{P}_{IFE_{2}} (F_{-1})$ for feeding into RedSCAB. Assuming we have $n$ RedSCABs, the $n$-th output $F_{n}$ can thus be written as:
\begin{equation} \label{eq:rdb}
    F_{n} = \mathcal{P}_{RedSCAB_{n}} (F_{n-1}) , 
\end{equation}
where $\mathcal{P}_{RedSCAB_{n}}$ represents the n-th RedSCAB operation ($n \geq 1$). Given the extracted local features from a set of RedSCAB, we apply our global feature fusion (GFF) to extract the global feature: 
\begin{equation}
    F_{GF} = \mathcal{P}_{GFF} (\{F_{1}, F_{2}, \dots, F_{n}\}) ,
\end{equation}
where $\{ \}$ means concatenation along feature channel and our global feature fusion function $\mathcal{P}_{GFF}$ consists of a $1 \times 1$ and $3 \times 3$ convolution layers to fuse the extracted local features from different levels of RedSCAB. The GFF output is used as input for our global residual learning:
\begin{equation}
    I = \mathcal{P}_{final} (F_{GF} + F_{-1}) ,
\end{equation}
The element-wise addition of global feature and initial feature are fed into our final $3 \times 3$ convolution layer for unregularized output.\\

\noindent\textbf{Residual Dense Spatial-Channel Block} contains four densely connected convolution layers, local feature fusion, local residual connection, and spatial-channel attention. In the $n$-th RedSCAB, the $t$-th convolution output is:
\begin{equation}
    F_{n}^{t} = \mathcal{H}_{n}^{t} \{F_{n-1}, F_{n}^{1}, \dots ,F_{n}^{t-1}\} ,
\end{equation}
where $\mathcal{H}_{n}^{t}$ denotes the $t$-th convolution followed by Leaky-ReLU in the $n$-th RedSCAB, $\{ \}$ means concatenation along feature channel, and the number of convolution $t \leq 4$. Then, we apply our local feature fusion (LFF), a $1 \times 1$ convolution layer, to fuse the output from the last RedSCAB and all convolution layers in current RedSCAB. Thus, the LFF output can be expressed as:
\begin{equation}
    F_{LF,n} = \mathcal{P}_{LFF,n} ( \{F_{n-1}, F_{n}^{1}, F_{n}^{2}, F_{n}^{3} ,F_{n}^{4}\} ) ,
\end{equation}
where $\mathcal{P}_{LFF,n}$ denotes the LFF operation. Then, it is fed into our Spatial-Channel Attention (SCA) module with two branches to re-weigh channel-wise features and spatial-wise features, as illustrated in Figure \ref{fig:rdscan}. The channel attention output $F_{CA,n}$ and spatial attention output $F_{SA,n}$ are fused together via $F_{SCA,n}=F_{CA,n}+F_{SA,n}$. Finally, we apply the local residual learning to SCA output by adding the residual connection from RedSCAB input, generating the $n$-th RedSCAB output:
\begin{equation}
    F_{n} = F_{SCA,n} + F_{n-1}
\end{equation}

\noindent\textbf{Spatial-Channel Attention} contains two Squeeze-and-Excitation branches for Channel Attention (CA) and Spatial Attention (SA), respectively \cite{hu2018squeeze,roy2018recalibrating}. Traditional CNNs treat channel-wise features and spatial-wise features equally. However, in an image reconstruction task, it is desirable to have the network focus more on informative features by acknowledging both the channel-wise feature interdependence and the spatial-wise contextual interdependence. The CA and SA structures are illustrated in orange and blue boxes in Figure \ref{fig:rdscan}, respectively. 

For CA, similar to \cite{hu2018squeeze}, we spatial-wise squeeze the input feature map using global average pooling, where the feature map is formulated as $F = [f_1, f_2, \dots, f_C]$ here with $f_n \in \mathbb{R}^{H \times W}$ denoting the individual feature channel. We flatten the global average pooling output, generating $v \in \mathbb{R}^{C}$ with its $z$-th element:
\begin{equation}
    v_z = \frac{1}{H \times W} \sum^H_i \sum^W_j f_z (i,j)
\end{equation}
where vector $v$ embeds the spatial-wise global information. Then, $v$ is fed into two fully connected layers with weights of $w_1 \in \mathbb{R}^{\frac{C}{2} \times C}$ and $w_2 \in \mathbb{R}^{C \times \frac{C}{2}}$, producing the channel-wise calibration vector:
\begin{equation}
    \hat{v} = \sigma(w_2 \eta(w_1 v)) 
\end{equation}
where $\eta$ and $\sigma$ are the ReLU and Sigmoid activation function, respectively. The calibration vector is applied to the input feature map using channel-wise multiplication:
\begin{equation}
    \hat{F}_{CA} = [f_1 \hat{v}_1, f_2 \hat{v}_2, \dots, f_C \hat{v}_C]
\end{equation}
where $\hat{v}_i$ indicates the importance of the $i$-th feature channel and lies in $[0,1]$. With CA embedded into our network, the calibration vector adaptively learns to emphasize the important feature channels while plays down the others. 

In SA, we formulate our feature map as $F = [f^{1,1}, \dots, f^{i,j}, \dots, f^{H,W}]$, where $f^{i,j} \in \mathbb{R}^{C}$ indicates the feature at spatial location $(i,j)$ with $i \in \{1,\dots,H\}$ and $j \in \{1,\dots,W\}$. We channel-wise squeeze the input feature map using a convolutional kernel with weights of $w_3 \in \mathbb{R}^{1 \times 1 \times C \times 1}$, generating a volume tensor $m = w_3 \circledast F$ with $m \in \mathbb{R}^{H \times W}$. Each $f^{i,j}$ is a linear combination of all feature channels at spatial location $(i,j)$. Then, the spatial-wise calibration volume that lies in $[0,1]$ can be written as:
\begin{equation}
    \hat{m} = \sigma(m) = \sigma(w_3 \circledast F)
\end{equation}
where $\sigma$ is the sigmoid activation function. Applying the calibration volume to the input feature map, we have:
\begin{equation}
    \hat{F}_{SA} = [f^{1,1} \hat{m}^{1,1}, \dots, f^{i,j} \hat{m}^{i,j}, \dots, f^{H,W} \hat{m}^{H,W}]
\end{equation}
where the calibration parameter $\hat{m}^{i,j}$ provides the relative importance of a spatial information of a given feature map. Similarly, with SA embedded into our network, the calibration volume learns to stress the most important spatial locations while ignores the irrelevant ones. 

Finally, channel-wise calibration and spatial-wise calibration are combined via element-wise addition operation $F_{SCA} = \hat{F}_{SA} + \hat{F}_{CA}$. With the two branch fusion, features at $(i,j,c)$ possess high activation only when they receive high activation from both SA and CA. Our SCA encourages the networks to re-calibrate the feature map such that more accurate and relevant feature maps can be learned.

\begin{algorithm2e*}[!htb]
\scriptsize
\caption{Recurrent Learning Framework}\label{alg:rec} 

\textbf{Input:} $\Pi$ = \{($I_{u_i}$, $S_{u_i}$, $I_{GT_i}$)\}, for $i \in \{1,\ldots,N\}$  \Comment*[r]{training dataset}

\textbf{Initialize:} $\theta_{init} :\gets \mathcal{N}(0,1)$  \Comment*[r]{initialize weights}

\For{$iter = 1$ to $K$} 
{

$T = \{(I_u, S_u, I_{GT})\} \gets \Pi$  \Comment*[r]{get training batch}

    \For{$j = 1$ to $Z$}
    {
    
    \eIf{$j=1$}{$input_j = I_u$ \;}{$input_j = output_{j-1}$ \;}
    
    $output_{net,j} \gets \mathcal{P}_{net}(input_j, \theta)$   \Comment*[r]{unregularized output}
    
    $output_{j} \gets \mathcal{P}_{SCL}(output_{net,j}, S_u, \lambda)$  \Comment*[r]{regularized output}
    
    }

$I_Z = output_Z$ \; 

$\theta :\gets min [\mathcal{L}(I_Z, X_{GT})]$  \Comment*[r]{optimization}

}

\textbf{Output} $\theta$  \Comment*[r]{return upon convergence}
\end{algorithm2e*} 

\subsection{Sinogram Consistency Layer}
\label{section:scl}
Our SCL aims to ensure the sinogram data fidelity that consists of three major operations: forward projection, sinogram consistency, and filtered back projection. Let $\varmathbb{F}$ and $\varmathbb{B}$ be forward projection and filtered back projection enabled by radon and inverse radon transform, respectively. The inputs to the SCL are the reconstructed image from the preceding RedSCAN ($I_{net}$) that is used to calculate $S_{net} = \varmathbb{F} I_{net} = \varmathbb{F} \mathcal{P}(I_{u};\theta)$, and $S_u = \varmathbb{F}_u I$ via forward projection. We can write a closed-form solution for the second term in (2) as:
\begin{equation}
  S_{rec}(i)=
  \begin{cases}
    \frac{\lambda S_{net}(i) + S_u(i)}{\lambda + 1} & \text{if $i \in \Omega$} \\[3pt]
    \quad\quad S_{net}(i) & \text{if $i \notin \Omega$}
  \end{cases}
\end{equation}
where $S_{rec}$ is the reconstructed sinogram, which is updated by the sinogram consistency constraint. Then, the image can be reconstructed via filtered back projection, that is, $I_{rec} = \varmathbb{B} S_{rec}$. To sum up, our SCL layer in the deep model can be written as:
\begin{equation}
    \mathcal{P}_{SCL}(I_{net}, S_u, \lambda) = \hat{R} (\mathcal{D} \varmathbb{F} I_{net} + \frac{1}{\lambda+1} S_u)
\end{equation}
where $\mathcal{D} = diag(e_1, e_2, \cdots, e_M)$ is a diagonal matrix with $e_i = \frac{\lambda}{1+\lambda}$ when $i \in \Omega$, and $e_i = 1$, $S_u=0$ when $i \notin \Omega$. To elaborate, when the $i$-th projection data is not acquired, our SCL directly estimates the $i$-th projection data from the projection data of the RedSCAN's output. Otherwise, the $i$-th projection data is a linear combination of the acquired projection data and projection data of the RedSCAN's output, regularized by noise level parameter $\lambda$. Assuming noiseless sinogram acquisition, i.e. $\lambda=0$, our SCL simply replaces the $i$-th predicted projection data by the acquired projection data. We empirically set $\lambda=0.001$, assuming low noise level.

\subsection{Recurrent Reconstruction Framework}
\label{section:rrf}
Our recurrent reconstruction framework with RedSCAN and SCL is summarized in Algorithm \ref{alg:rec}. The input of R$^2$edSCAN is the initial FBP reconstruction ($I_u$) based on limited view sinogram data ($S_u$). In each recurrent block, the input first passes through RedSCAN for generating unregularized reconstruction and then feed into SCL for generating regularized reconstruction output. Given the output from the last/$Z$-th recurrent block and ground truth FBP reconstruction ($I_{GT}$) from fully sampled sinogram, we train the R$^2$edSCAN using the following loss function: 
\begin{equation}
    \mathcal{L}_1 = ||I_Z - I_{GT}||_1
\end{equation}
The number of recurrent blocks is set to 4 in our experiments.

\begin{figure*}[htb!]
\centering
\includegraphics[width=1.00\textwidth]{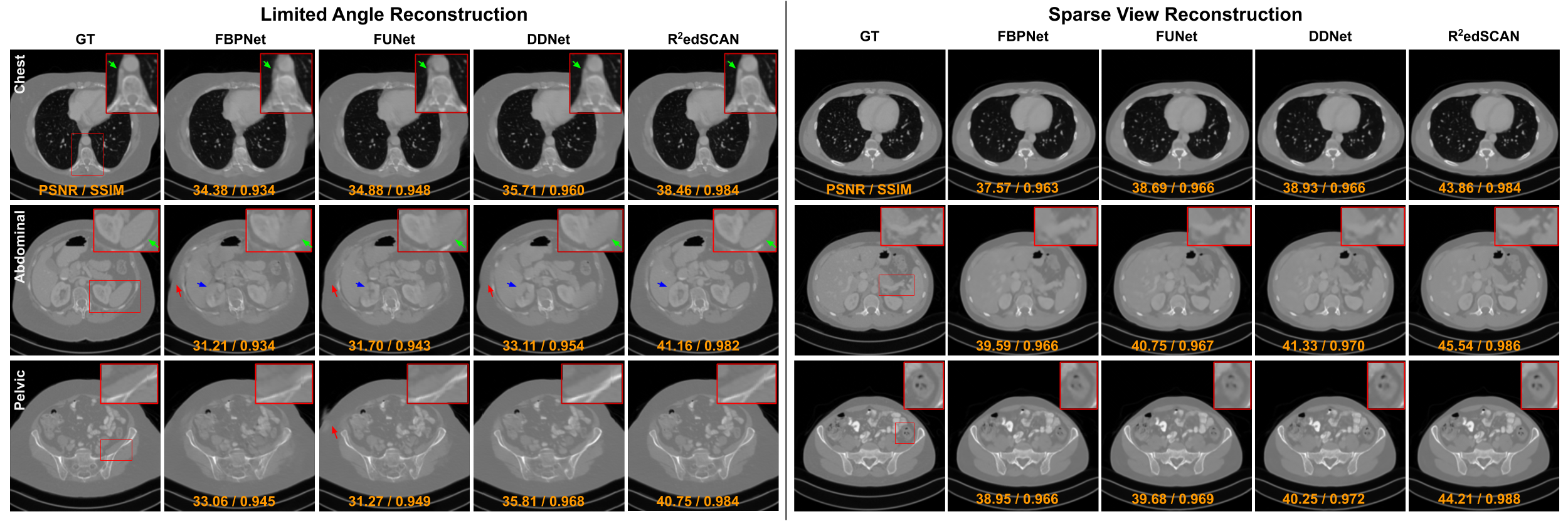}
\caption{Comparison of \textbf{limited angle reconstructions} and \textbf{sparse view reconstructions} in chest, abdominal, and pelvic CT scans. In our LA chest reconstruction, important arterial structure (green arrows) is better preserved using our R$^2$edSCAN. Similarly for spleen (green arrows) and kidney boundary (blue arrows) in the abdominal reconstruction. In addition, patient's boundary artifacts (red arrows) are significantly suppressed. The PSNR and SSIM values are indicated at the bottom. Note that GT is the ground truth reconstruction for comparison.}
\label{fig:comp_la_sv_aapm}
\end{figure*}


\begin{table*} [htb!]
\scriptsize
\centering
\caption{Quantitative comparison of \textbf{limited angle reconstruction} and \textbf{sparse view reconstruction} results using PSNR (dB) and SSIM. Best and second best results are marked in \textcolor{red}{red} and \textcolor{blue}{blue}, respectively.}
\label{tab:PSNRandSSIM_AAPM}
    \begin{tabular}{l|c|c|c|c||c|c|c|c|c}
        \hline
        \multirow{2}{*}{\textbf{PSNR/SSIM}}  & \multicolumn{4}{c}{\textbf{Limited Angle Reconstruction}} &  \multicolumn{4}{c}{\textbf{Sparse View Reconstruction}}  & \textbf{Time} \Tstrut\Bstrut\\
        \cline{2-9}
                                             & Chest       & Abdominal   & Pelvic      & All              & Chest       & Abdominal   & Pelvic      & All               & \textbf{(ms)} \Tstrut\Bstrut\\
        \hline
        FBP                                  & 16.73/.401  & 16.63/.441  & 17.32/.401  & 17.04/.413       & 25.68/.498  & 27.24/.548  & 26.40/.488  & 26.57/.507        & 2.21  \Tstrut\Bstrut\\
        \hline
        FBPNet \cite{jin2017deep}            & 30.37/.943  & 32.61/.944  & 33.71/.954  & 33.02/.950       & 37.27/.959  & 38.55/.960  & 38.76/.964  & 38.54/.962        & 7.23  \Tstrut\Bstrut\\
        \hline
        DDNet \cite{zhang2018sparse}         & 33.12/.963  & 34.88/.961  & 36.33/.969  & 35.55/.966       & 39.45/.966  & 40.66/.967  & 40.62/.970  & 40.50/.969        & 5.03  \Tstrut\Bstrut\\
        \hline
        FUNet \cite{han2018framing}          & 31.11/.950  & 33.15/.952  & 34.61/.959  & 33.80/.956       & 39.18/.963  & 40.21/.964  & 40.06/.965  & 40.01/.965        & 5.65  \Tstrut\Bstrut\\
        \hline
        RedSCAN                              & \textcolor{blue}{33.78}/\textcolor{blue}{.962}  & \textcolor{blue}{35.13}/\textcolor{blue}{.965}  & \textcolor{blue}{36.35}/\textcolor{blue}{.970}  & \textcolor{blue}{35.82}/\textcolor{blue}{.969}       & \textcolor{blue}{41.03}/\textcolor{blue}{.972}  & \textcolor{blue}{42.10}/\textcolor{blue}{.971}  & \textcolor{blue}{41.95}/\textcolor{blue}{.974}  & \textcolor{blue}{41.98}/\textcolor{blue}{.973}        & 5.02  \Tstrut\Bstrut\\
        \hline
        \hline
        R-UNet                               & 32.73/.952  & 35.57/.954  & 35.99/.962  & 35.51/.959       & 38.68/.966  & 39.95/.969  & 40.13/.973  & 39.92/.971        & 117.1  \Tstrut\Bstrut\\
        \hline
        R$^2$edSCAN                            & \textcolor{red}{38.84}/\textcolor{red}{.987}  & \textcolor{red}{40.58}/\textcolor{red}{.986}  & \textcolor{red}{41.03}/\textcolor{red}{.989}  & \textcolor{red}{40.75}/\textcolor{red}{.987}       & \textcolor{red}{43.52}/\textcolor{red}{.988}  & \textcolor{red}{44.50}/\textcolor{red}{.988}  & \textcolor{red}{44.53}/\textcolor{red}{.990}  & \textcolor{red}{44.41}/\textcolor{red}{.989}        & 119.9  \Tstrut\Bstrut\\
        \hline
    \end{tabular}
\end{table*}

\section{EXPERIMENTS AND RESULTS}
\subsection{Data Preparation and Training}
We used two large-scale dataset for our experiments. In our first dataset, we collected 10 whole body CT scans from the AAPM Low Dose CT Grand Challenge \cite{mccollough2016tu}. Each 3D scan contains 318 $\sim$ 856 2D slices covering a range of anatomical regions from chest to abdomen to pelvis. From the AAPM dataset, the 2D dataset of 3397 images without lesion are split patient-wise into 1834 training images, 428 validation images, and 1135 test images. To evaluate the reconstruction performance on CT image with important pathological findings, in our second dataset, we collected 2900 2D CT slices from the DeepLesion dataset \cite{yan2018deeplesion}, which consists of 8 different lesion types (bone:240, liver:380, lung:380, kidney:380, mediastinum:380, abdominal:380, pelvis:380, soft-tissue:380). We split the DeepLesion 2D dataset into 1960 training images (110 for bone, 250 for each of the rest lesion types), 300 validation images (50 slices for each lesion types), 640 test images (80 slices for each lesion types). We combined two dataset for training and testing.

For each image, the fully sampled sinogram data $S$ was generated via 240 simulated parallel-beam projection views covering 0 to 180 degrees. In sparse view experiments, we uniformly sampled 40 projection views from the 240 projection views to form $S_u$, mimicking 6 fold radiation dose reduction. In limited angle experiments, we sampled 160 (out of the 240 total) projection views that lies within 0 to 120 degrees for our $S_u$. The reconstructed image $I$ and $I_u$ were obtained by applying FBP with ramp filter to $S$ and $S_u$, respectively.

We implemented our R$^2$edSCAN in Pytorch\footnote{http://pytorch.org/}, and trained it on an NVIDIA Quadro RTX 8000 GPU with 48G memory. The Adam solver \cite{kingma2014adam} was used to optimize our models with a momentum of 0.99 and a 0.0005 learning rate. We used a batch size of 4 during training.

\begin{figure}[htb!]
\centering
\includegraphics[width=0.46\textwidth]{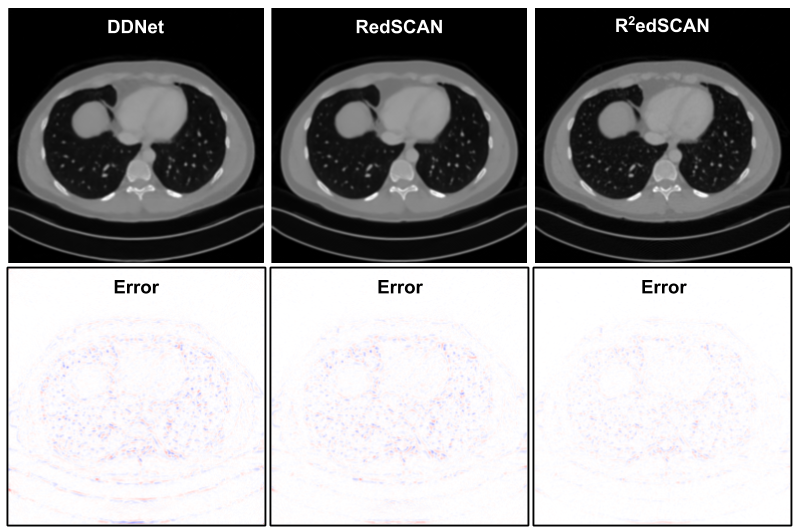}
\caption{Illustration of sparse view RedSCAN reconstruction. RedSCAN can provide better reconstruction over DDNet with less errors. Embedding our RedSCAN into our recurrent reconstruction framework (R$^2$edSCAN) further enhances the reconstruction.}
\label{fig:comp_r}
\end{figure}

\begin{figure*}[htb!]
\centering
\includegraphics[width=1.00\textwidth]{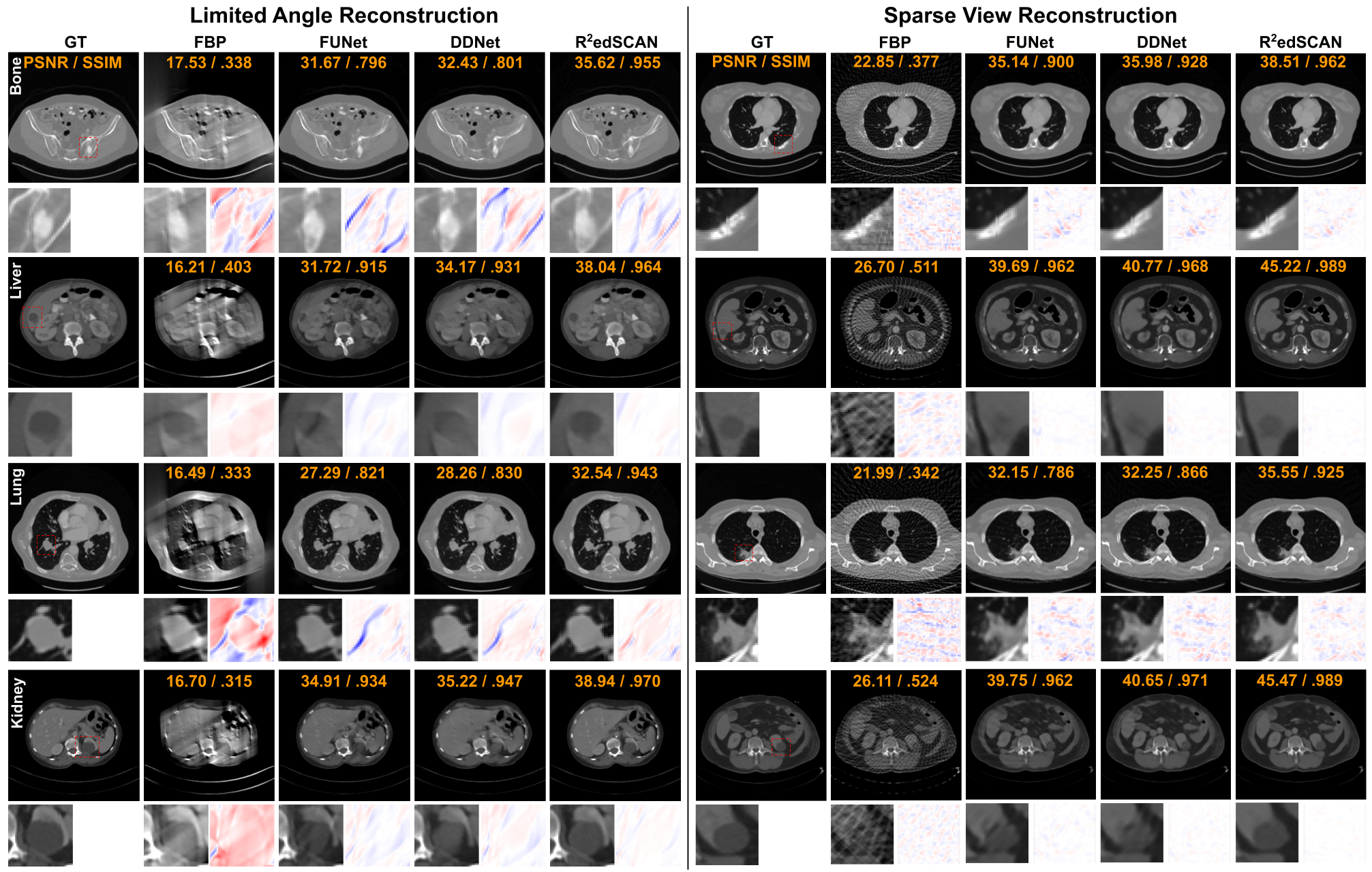}
\caption{Comparison of \textbf{limited angle reconstructions} and \textbf{sparse view reconstructions} in CT scans with lesions. The lesion region zoom-in view and the corresponding difference to ground truth are shown on the bottom row. The PSNR and SSIM values are indicated at the top left.}
\label{fig:comp_la_sv_deeplesion}
\end{figure*}

\subsection{Experimental Results}
For quantitative evaluation, both SV and LA results were evaluated using Peak Signal-to-Noise Ratio (PSNR) and Structural Similarity Index (SSIM) by comparing the synthetic SV and LV reconstructions to the ground truth reconstruction from FBP of fully sampled sinogram. For comparative study, we compared our results on both tasks against several existing limited view reconstruction methods, including the combination of Densenet and Deconvolution (DDNet) \cite{zhang2018sparse}, Framing UNet (FUNet) \cite{han2018framing}, and FBPNet \cite{jin2017deep}.

The qualitative comparison of different limited angle reconstruction methods with AAPM dataset is shown in Figure \ref{fig:comp_la_sv_aapm}. As we can observe in chest region, previous methods have difficulties in reconstructing small anatomical structure, i.e. arteries. Similarly, with crowded organs in abdominal region, the organ boundaries are challenging to recover by previous methods along with additional patient boundary artifacts. Our R$^2$edSCAN with advanced network design and sinogram consistency constraint can provide superior limited angle reconstruction in terms of organ boundary recovery, small structure recovery, and boundary artifact elimination. Table \ref{tab:PSNRandSSIM_AAPM} outlines the quantitative comparison of different methods on limited angle reconstruction with AAPM dataset. Our R$^2$edSCAN significantly outperforms previous state-of-the-art methods. Compared to the best previous method's performance of DDNet \cite{zhang2018sparse}, we improve PSNR from $35.55$ to $40.75$ and SSIM from $0.966$ to $0.987$, respectively.

The qualitative comparison of different sparse view reconstruction methods with AAPM datset is shown in Figure \ref{fig:comp_la_sv_aapm}. Similar to the observations from limited angle experiments above, our R$^2$edSCAN yields high-quality reconstruction in crowded soft tissue area with fine details, i.e., splenic vein (2nd row) and small bowel (3rd row). As evidenced in Table \ref{tab:PSNRandSSIM_AAPM}, our R$^2$edSCAN achieves the best results among various previous methods. Compared to the best previous method's performance of DDNet \cite{zhang2018sparse}, we improve PSNR from $40.50$ to $44.41$ and SSIM from $0.969$ to $0.989$, respectively. It is also worth noticing in Table \ref{tab:PSNRandSSIM_AAPM} that our RedSCAN without SCL obtains the second best result for both LA and SV tasks, demonstrating our efficient network design in image restoration. Figure \ref{fig:comp_r} demonstrates an example of our RedSCAN sparse view reconstruction and the corresponding error image.

\begin{table*} [htb!]
\scriptsize
\centering
\caption{Quantitative comparison of \textbf{limited angle reconstruction} results using PSNR (dB) and SSIM. Best results are marked in \textcolor{red}{red}. NF denotes no founding CT images from AAPM dataset.}
\label{tab:PSNRandSSIM_DEEPLESION_LA}
    \begin{tabular}{|l||c|c|c|c|c|c|c|c|c||c|}
        \hline
        \textbf{PSNR/SSIM}           & Bone        & Abdomen     & Mediastinum & Liver       & Lung        & Kidney      & Soft Tissue & Pelvis      & NF          & \textbf{All}         \Tstrut\Bstrut\\
        \hline
        FBP                          & 17.09/.337  & 16.82/.383  & 17.30/.334  & 16.71/.363  & 17.20/.313  & 16.95/.391  & 17.31/.331  & 17.21/.360  & 17.04/.413  & 17.06/.380  \Tstrut\Bstrut\\
        FBPNet\cite{jin2017deep}     & 31.29/.860  & 33.53/.928  & 30.14/.908  & 32.11/.918  & 29.50/.822  & 33.74/.928  & 31.63/.909  & 32.80/.928  & 33.02/.950  & 32.39/.923  \Tstrut\Bstrut\\
        DDNet\cite{zhang2018sparse}  & 33.16/.888  & 35.46/.946  & 32.08/.929  & 34.04/.940  & 31.46/.859  & 36.11/.947  & 33.66/.927  & 34.70/.941  & 35.55/.966  & 34.63/.943  \Tstrut\Bstrut\\
        FUNet\cite{han2018framing}   & 31.71/.878  & 33.99/.937  & 30.32/.916  & 32.64/.928  & 29.90/.839  & 34.29/.937  & 32.00/.914  & 33.06/.931  & 33.80/.956  & 32.96/.932  \Tstrut\Bstrut\\
        RedSCAN                      & 33.70/.927  & 36.59/.950  & 32.63/.930  & 34.81/.943  & 31.77/.910  & 37.49/.952  & 35.12/.939  & 35.72/.951  & 35.82/.969  & 35.21/.949  \Tstrut\Bstrut\\
        \hline
        R-UNet                       & 33.63/.924  & 36.51/.948  & 32.48/.926  & 34.63/.939  & 31.48/.908  & 37.06/.950  & 34.74/.935  & 35.65/.946  & 35.51/.959  & 34.98/.945  \Tstrut\Bstrut\\
        R$^2$edSCAN                    & \textcolor{red}{36.89}/\textcolor{red}{.959}  & \textcolor{red}{40.29}/\textcolor{red}{.978}  & \textcolor{red}{36.03}/\textcolor{red}{.963}  & \textcolor{red}{38.32}/\textcolor{red}{.972}  & \textcolor{red}{34.28}/\textcolor{red}{.949}  & \textcolor{red}{41.30}/\textcolor{red}{.982}  & \textcolor{red}{38.13}/\textcolor{red}{.970}  & \textcolor{red}{39.30}/\textcolor{red}{.978}  & \textcolor{red}{40.34}/\textcolor{red}{.982}  & \textcolor{red}{39.12}/\textcolor{red}{.975}  \Tstrut\Bstrut\\
        \hline
    \end{tabular}
\end{table*}

\begin{table*} [htb!]
\scriptsize
\centering
\caption{Quantitative comparison of \textbf{sparse view reconstruction} results using PSNR (dB) and SSIM. Best results are marked in \textcolor{red}{red}. NF denotes no founding CT images from AAPM dataset.}
\label{tab:PSNRandSSIM_DEEPLESION_SV}
    \begin{tabular}{|l||c|c|c|c|c|c|c|c|c||c|}
        \hline
        \textbf{PSNR/SSIM}           & Bone        & Abdomen     & Mediastinum & Liver       & Lung        & Kidney      & Soft Tissue & Pelvis      & NF          & \textbf{All}         \Tstrut\Bstrut\\
        \hline
        FBP                          & 24.47/.434  & 26.31/.514  & 25.19/.443  & 26.17/.502  & 23.03/.390  & 26.58/.515  & 25.84/.460  & 26.08/.484  & 26.57/.507  & 25.98/.486  \Tstrut\Bstrut\\
        FBPNet\cite{jin2017deep}     & 35.76/.914  & 38.44/.959  & 36.87/.947  & 38.27/.959  & 34.20/.884  & 38.57/.962  & 37.67/.950  & 38.48/.963  & 38.54/.962  & 37.87/.951  \Tstrut\Bstrut\\
        DDNet\cite{zhang2018sparse}  & 37.97/.940  & 40.94/.965  & 39.01/.955  & 40.76/.964  & 35.97/.918  & 40.95/.966  & 40.35/.960  & 41.08/.966  & 40.50/.968  & 40.04/.961  \Tstrut\Bstrut\\
        FUNet\cite{han2018framing}   & 36.90/.911  & 39.87/.958  & 38.09/.948  & 39.71/.958  & 35.21/.886  & 39.96/.958  & 39.30/.952  & 40.05/.961  & 40.01/.965  & 39.28/.952  \Tstrut\Bstrut\\
        RedSCAN                      & 38.98/.954  & 42.22/.977  & 40.01/.970  & 42.02/.977  & 36.70/.933  & 42.40/.979  & 41.53/.975  & 42.30/.980  & 41.65/.973  & 41.18/.970  \Tstrut\Bstrut\\
        \hline
        R-UNet                       & 36.15/.918  & 38.94/.953  & 37.17/.949  & 38.67/.955  & 34.27/.887  & 39.07/.965  & 38.08/.951  & 39.09/.964  & 39.92/.971  & 38.72/.956  \Tstrut\Bstrut\\
        R$^2$edSCAN                    & \textcolor{red}{41.24}/\textcolor{red}{.972}  & \textcolor{red}{45.04}/\textcolor{red}{.988}  & \textcolor{red}{42.55}/\textcolor{red}{.983}  & \textcolor{red}{44.77}/\textcolor{red}{.988}  & \textcolor{red}{38.98}/\textcolor{red}{.960}  & \textcolor{red}{45.29}/\textcolor{red}{.989}  & \textcolor{red}{44.22}/\textcolor{red}{.986}  & \textcolor{red}{39.30}/\textcolor{red}{.978}  & \textcolor{red}{44.87}/\textcolor{red}{.989}  & \textcolor{red}{43.81}/\textcolor{red}{.983}  \Tstrut\Bstrut\\
        \hline
    \end{tabular}
\end{table*}

\begin{table*} [htb!]
\scriptsize
\centering
\caption{Quantitative comparison of \textbf{limited angle reconstruction} and \textbf{sparse view reconstruction} in lesion regions using PSNR (dB) and SSIM. Best results are marked in \textcolor{red}{red}.}
\label{tab:PSNRandSSIM_DEEPLESION_ZOOM}
    \begin{tabular}{|l||c|c|c|c|c|c|c|c|}
        \hline
        \textbf{PSNR/SSIM - LA}      & Bone        & Abdomen     & Mediastinum & Liver       & Lung        & Kidney      & Soft Tissue & Pelvis      \Tstrut\Bstrut\\
        \hline
        FBP                          & 19.01/.691  & 27.21/.876  & 18.01/.663  & 26.87/.891  & 18.03/.608  & 26.77/.889  & 20.38/.766  & 21.68/.815  \Tstrut\Bstrut\\
        FBPNet\cite{jin2017deep}     & 26.53/.843  & 32.93/.922  & 26.88/.873  & 34.02/.949  & 26.95/.863  & 33.14/.920  & 32.85/.925  & 30.59/.887  \Tstrut\Bstrut\\
        DDNet\cite{zhang2018sparse}  & 28.87/.894  & 35.36/.949  & 29.22/.916  & 36.01/.965  & 29.32/.909  & 35.38/.941  & 35.29/.947  & 33.26/.923  \Tstrut\Bstrut\\
        FUNet\cite{han2018framing}   & 26.94/.860  & 33.76/.936  & 27.94/.899  & 34.69/.958  & 27.56/.879  & 33.56/.927  & 33.38/.933  & 31.38/.906  \Tstrut\Bstrut\\
        R$^2$edSCAN                    & \textcolor{red}{31.42}/\textcolor{red}{.923}  & \textcolor{red}{39.92}/\textcolor{red}{.971}  & \textcolor{red}{34.24}/\textcolor{red}{.960}  & \textcolor{red}{40.35}/\textcolor{red}{.977}  & \textcolor{red}{32.19}/\textcolor{red}{.942}  & \textcolor{red}{40.20}/\textcolor{red}{.968}  & \textcolor{red}{40.22}/\textcolor{red}{.971}  & \textcolor{red}{36.92}/\textcolor{red}{.953}  \Tstrut\Bstrut\\
        \hline
        \textbf{PSNR/SSIM - SV}      & Bone        & Abdomen     & Mediastinum & Liver       & Lung        & Kidney      & Soft Tissue & Pelvis      \Tstrut\Bstrut\\
        \hline
        FBP                          & 30.14/.809  & 33.57/.777  & 31.46/.814  & 32.41/.709  & 27.49/.707  & 33.27/.773  & 30.52/.689  & 34.21/.819  \Tstrut\Bstrut\\
        FBPNet\cite{jin2017deep}     & 33.86/.930  & 39.87/.949  & 34.87/.940  & 42.70/.972  & 32.30/.900  & 40.11/.949  & 38.69/.947  & 39.36/.953  \Tstrut\Bstrut\\
        DDNet\cite{zhang2018sparse}  & 34.50/.939  & 40.81/.948  & 35.44/.956  & 43.63/.977  & 33.29/.911  & 40.96/.949  & 39.94/.957  & 40.51/.962  \Tstrut\Bstrut\\
        FUNet\cite{han2018framing}   & 33.64/.930  & 39.70/.957  & 36.60/.944  & 42.54/.974  & 32.78/.908  & 39.92/.956  & 39.06/.949  & 39.12/.951  \Tstrut\Bstrut\\
        R$^2$edSCAN                    & \textcolor{red}{36.81}/\textcolor{red}{.962}  & \textcolor{red}{43.92}/\textcolor{red}{.978}  & \textcolor{red}{39.31}/\textcolor{red}{.976}  & \textcolor{red}{46.66}/\textcolor{red}{.988}  & \textcolor{red}{35.90}/\textcolor{red}{.949}  & \textcolor{red}{44.21}/\textcolor{red}{.977}  & \textcolor{red}{43.68}/\textcolor{red}{.980}  & \textcolor{red}{43.09}/\textcolor{red}{.978}  \Tstrut\Bstrut\\
        \hline
    \end{tabular}
\end{table*}

As CT scan is often used for disease diagnosis, we also evaluated the reconstruction performance on CT images with 8 different lesion types. Figure \ref{fig:comp_la_sv_deeplesion} illustrates the qualitative comparison of various limited angle and sparse view reconstruction methods on 4 major lesion types. As we can observe, the liver lesion and kidney lesion are hard to recover by previous methods because these lesions have low contrast to the soft-tissue background, and their visualization are further degraded by the limited angle artifacts. Similarly, the bone lesion and lung lesion are also challenging to recover by previous methods due to their complex lesion texture and surrounding tissue structures. However, our R$^2$edSCAN can provide superior recovery of the shape and texture of the lesion even under these difficult conditions. For example, our liver and kidney reconstructions on the last column can provide clear lesion boundary which is critical for lesion progression assessment. The lung bronchi that originally diminished on FBP reconstruction can also be recovered by our R$^2$edSCAN. Table \ref{tab:PSNRandSSIM_DEEPLESION_LA} and \ref{tab:PSNRandSSIM_DEEPLESION_SV} summarizes the reconstruction performance on CT images with 8 different lesion types. Our R$^2$edSCAN achieves mean SSIM over 0.97 on limited angle reconstruction and mean SSIM over 0.98 on sparse view reconstruction which consistently outperforms previous reconstruction methods on all 8 lesion types. Furthermore, Table \ref{tab:PSNRandSSIM_DEEPLESION_ZOOM} outlines the reconstruction performance in the cropped lesion regions. Our R$^2$edSCAN obtains the best lesion region reconstructions across all lesion types. 

\subsection{Ablation Studies} 
\begin{figure}[htb!]
\centering
\includegraphics[width=0.49\textwidth]{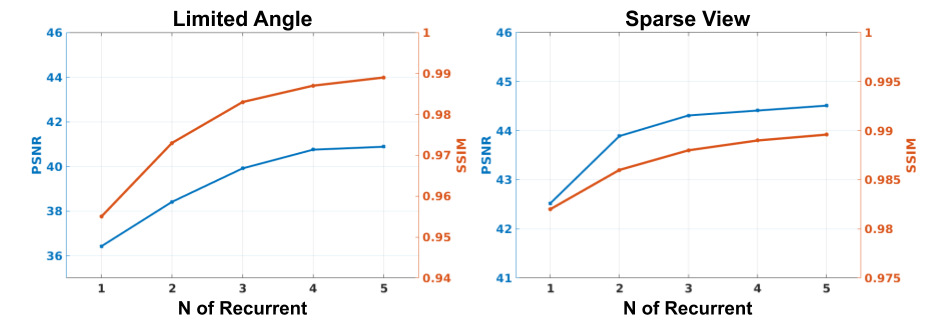}
\caption{The effect of increasing the number of recurrent blocks (Z) in our R$^2$edSCAN for sparse view reconstruction (right) and limited angle reconstruction (left). Reconstruction quality gradually increases as the number of recurrent blocks increases. }
\label{fig:effect_rec}
\end{figure}

We analyzed the effect of increasing the number of recurrent blocks in our R$^2$edSCAN. The result is summarized in Figure \ref{fig:effect_rec} and evaluated using AAPM dataset. As we can observe, using more recurrent blocks boosts the reconstruction performance, while the rate of improvement starts to converge after the number of recurrent blocks reaches 3. 

We analyzed the effect of different attention mechanisms used in our R$^2$edSCAN. The result is illustrated in Table \ref{tab:att_comp} and evaluated using AAPM dataset. As we can observe, both channel attention and spatial attention can improve the reconstruction performance, and the combination of both attentions provides the best performance with the least variation.

\begin{table} [htb!]
\scriptsize
\centering
\caption{Attention mechanism analysis using PSNR and SSIM. \textcolor{green}{\cmark} and \xmark\space means Channel Attention (CA) and Spatial Attention (SA) used and not used in our R$^2$edSCAN. The optimal results are in bold.}
\label{tab:att_comp}
\begin{tabular}{l|c c|c c}
    \hline
    Task                                             & CA                           & SA                          & PSNR                                & SSIM                                \Tstrut\Bstrut\\
    \hline   
    \multirow{4}{*}{Limited Angle}                   & \xmark                       & \xmark                      & $40.34 \pm 2.08 $                   & $0.983 \pm 0.028 $                    \Tstrut\Bstrut\\
                                                     & \textcolor{green}{\cmark}    & \xmark                      & $40.58 \pm 2.02 $                   & $0.985 \pm 0.021 $                    \Tstrut\Bstrut\\
                                                     & \xmark                       & \textcolor{green}{\cmark}   & $40.42 \pm 2.05 $                   & $0.984 \pm 0.025 $                    \Tstrut\Bstrut\\
                                                     & \textcolor{green}{\cmark}    & \textcolor{green}{\cmark}   & $\mathbf{40.75 \pm 1.98}$           & $\mathbf{0.987 \pm 0.020}$                    \Tstrut\Bstrut\\
    \hline            
    \multirow{4}{*}{Sparse View}                     & \xmark                       & \xmark                      & $44.32 \pm 1.76 $                   & $0.985 \pm 0.023 $                    \Tstrut\Bstrut\\
                                                     & \textcolor{green}{\cmark}    & \xmark                      & $44.39 \pm 1.68 $                   & $0.987 \pm 0.019 $                    \Tstrut\Bstrut\\
                                                     & \xmark                       & \textcolor{green}{\cmark}   & $44.35 \pm 1.70 $                   & $0.986 \pm 0.021 $                    \Tstrut\Bstrut\\
                                                     & \textcolor{green}{\cmark}    & \textcolor{green}{\cmark}   & $\mathbf{44.41 \pm 1.69}$           & $\mathbf{0.989 \pm 0.017}$                    \Tstrut\Bstrut\\
    \hline             
\end{tabular}
\end{table}

\section{DISCUSSION}
In this paper, a novel reconstruction framework, named R$^2$edSCAN, is proposed. Inspired by the recent advances in image super-resolution network designs and the sinogram consistency constraint in MBIR, we designed a customized RedSCAN as our backbone image reconstruction network, and we built a sinogram consistency layer that can be embedded in deep networks. First of all, our RedSCAN is developed based on image super-resolution network \cite{zhang2020residual} with an addition of spatial-channel attention, which allows our RedSCAN to re-calibrate the channel attention and gives different levels of attention on recovering texture details at different spatial locations, as artifact distribution is not uniform in the image. Then, we develop SCL that can be concatenated to the RedSCAN's recurrent outputs to ensure the sinogram consistency at the sampled projection views. Our SCL based on the fast analytical FBP solution allows it to be embedded in deep network and used during training and inference. 

We demonstrate the feasibility of our R$^2$edSCAN on both LA and SV tomographic reconstruction tasks, as shown in the result section. Firstly, the LA acquisition is more difficult to reconstruct as compared to the SV acquisition since a range of projection angles are not covered in the LA acquisition. Severe image artifacts at these projection angles can be observed when using conventional FBP. As a result, the general performance of LA reconstructions are inferior to the SV reconstruction performance. For LA reconstruction, while previous methods can mitigate the artifacts and recover PSNR up to $35.55$ and SSIM up to $0.966$, they still have difficulties in recovering the organ boundaries that are critical for clinical diagnosis and treatment planning. Our R$^2$edSCAN provides superior reconstructions with clear organ boundaries and is able to improve the PSNR to $40.75$ and SSIM to $0.987$. For SV reconstruction, while previous methods can generate visually plausible image content, the reconstruction prediction without sinogram consistency constraint can result in artificial texture which is undesirable in clinical tasks. Our R$^2$edSCAN with SCL can better preserve the image fidelity by incorporating the already-sampled projection data, resulting in best PSNR and SSIM performance.

Furthermore, we demonstrate the feasibility of our R$^2$edSCAN on CT lesion imaging under LA and SV conditions. Lesion is highly heterogeneous, and CT is one of the primary tool for diagnosis. Obtaining high-quality lesion region reconstruction under LA and SV is essential for disease diagnosis, staging, as well as planning and evaluation of treatment. While previous methods can reduce the reconstruction artifacts from the whole image perspective, the reconstruction in lesion region with high heterogeneity is still unsatisfying - the lesion boundary and texture are highly distorted by previous methods which will negatively impact the subsequent treatment options. On the other hand, our R$^2$edSCAN with sinogram consistency constraint can better preserve the lesion reconstruction even the lesions are highly heterogeneous. For example, the supplying vessels of LA lung lesion in Figure \ref{fig:comp_la_sv_deeplesion} are totally missed by previous methods, while our R$^2$edSCAN can correctly recover it. The complex interior texture of SV lung lesion in Figure \ref{fig:comp_la_sv_deeplesion} is highly distorted by previous methods, but our R$^2$edSCAN can still preserve the structure. In Figure \ref{fig:comp_la_sv_deeplesion}, liver and kidney lesions embedded in soft-tissue background with low contrast are prone to smooth-out in SV and distorted in LA by previous methods, and our R$^2$edSCAN can recover clear boundary and contrast of the lesions.

The presented work also has potential limitations. First of all, the inference time is longer compared to the previous deep learning based methods, as illustrated in Table \ref{tab:PSNRandSSIM_AAPM}. This is caused by the recurrent framework design with SCL layer interleaved. On one hand, the iterative reconstruction prediction will increase the computation time. On the other hand, even though FBP is a fast analytic solution, the forward projection and FBP operations in SCL still consume computation times. The combination of these two results in longer training and inference time. However, the inference time is about $120$ ms which is acceptable and much faster than previous MBIR methods. Secondly, while increasing the number of recurrent iterations in R$^2$edSCAN improves the performance, the memory consumption will increase along with longer training and inference time. As illustrated in Figure \ref{fig:effect_rec}, the increase in performance starts to converge after $n=3$. Thus, in this work, we set $n=4$ to balance the memory consumption and inference time of our R$^2$edSCAN. Thirdly, our current SCL focuses on a parallel-beam imaging geometry, which is sufficient to demonstrate the feasibility of our sinogram consistency layer. While other projection geometries, such as fan/cone beam, are not illustrated in our current work, we believe a SCL adaptive for the other geometry can be deduced similarly as that for the parallel-beam geometry, which will be a topic for our future work.

The architecture of our R$^2$edSCAN also suggests several interesting topics for future studies. The first one is combining the sinogram consistency layer with the deep learning based radon inversion techniques \cite{he2020radon}. The recurrent reconstruction framework with sinogram consistency can provide the projection domain constraint during the radon inversion via deep learning. It can potentially improve the inversion stability, yielding reconstruction with better data fidelity. 

Secondly, given the superior lesion region reconstruction performance demonstrated in the result sections, our framework could also potentially improve the projection data based Computer-Aided Diagnosis (CAD). Recently, there are increasing interests on combining limited-view reconstruction and CAD for a joint reconstruction-CAD network structure, and improved CAD performance is expected with such an end-to-end training strategy \cite{wei2018joint,adler2018task}. We believe that our R$^2$edSCAN with high-quality lesion region reconstruction would provide new opportunities for these kinds of studies. 

Lastly, CT metal artifact reduction (MAR) under limited-view acquisition is an important research direction. Current MAR techniques are mostly limited to full-view acquisition \cite{lin2019dudonet,katsura2018current}. The current state-of-the-art metal artifact reduction algorithm, such as DuDoNet \cite{lin2019dudonet}, utilizes projection space and image space simultaneously which is similar to our R$^2$edSCAN design. Our R$^2$edSCAN could potentially integrated with current MAR network for MAR under limited view conditions. 

\section{CONCLUSION}
In this work, we proposed a recurrent reconstruction framework with RedSCAN and sinogram consistency layer, a novel framework for limited view tomographic reconstruction. The proposed sinogram consistency layer is interleaved in our recurrent framework to ensure the sampled sinogram is consistent in sinogram domain with the network recurrent output. A customized image restoration network is used as the backbone in the recurrent framework. Comprehensive evaluation demonstrates that our R$^2$edSCAN can outperform existing methods on both limited angle and sparse view reconstruction tasks, providing high-quality tomographic reconstruction while reducing radiation dose and shortening scanning time.

\section{ACKNOWLEDGEMENT}
This work was supported by funding from the National Institutes of Health (NIH) under grant number R01EB025468. BZ was supported by the Biomedical Engineering Ph.D. fellowship from Yale University.

\bibliographystyle{IEEEtran}
\bibliography{bibliography}

\end{document}